\documentclass{llncs}

\usepackage{makeidx}
\usepackage{amsfonts}
\usepackage{amsmath}
\usepackage{graphicx}
\usepackage{comment}

\def\XR#1{\xrightarrow[\rule{2cm}{0pt}]{#1}}
\def\XL#1{\xleftarrow[\rule{2cm}{0pt}]{#1}}
\def\Z{\mathbb Z}

\begin{document}
\frontmatter
\pagestyle{headings}
\mainmatter

\title{The problem of popular primes: Logjam}

\author{Wouter Bokslag}
\institute{Technische Universiteit Eindhoven}

\maketitle

\begin{abstract}
This paper will discuss the Logjam attack on TLS. The Logjam attack allows, under certain conditions, to defeat the security provided by TLS. This is done by manipulating server and client into using weak and deprecated \emph{export grade} crypto, and subsequently breaking the Diffie-Hellman key exchange. We explore how the attack works conceptually and how exactly TLS is vulnerable to this attack. Also, the conditions under which the attack can be mounted are discussed, and an estimate of the impact of the attack is presented. Lastly, several mitigations are presented. 
\end{abstract}

\section{Introduction}
In recent days, awareness about the security of communications is on the rise. More and more websites provide secure connections to protect visitors from Man-in-the-Middle attacks that might compromise confidentiality and/or integrity of the communication. Initiatives like Let's Encrypt\cite{letsencrypt} further facilitate the transition towards encrypted communication.

TLS is a cryptographic protocol that facilitates setting up and using secure communication channels, supporting several authentication methods and different cryptographic suites. TLS is well known for its use in HTTPS, but also IMAP, SMTP and many other protocols rely on TLS. As such, the security of TLS is very important. In the past, various vulnerabilities in the TLS protocol or in its implementations were found, such as the Lucky 13\cite{al2013lucky} and FREAK\cite{smacktls}\cite{beurdouche2015messy} attacks. 

In this paper, we discuss the Logjam attack, that targets the Diffie-Hellman key exchange protocol. In Section \ref{sec:diffiehellman} we will discuss Diffie-Hellman, and briefly cover how it can be attacked. The parts of the TLS protocol that are relevant are discussed in Section \ref{sec:tls}, while the actual Logjam attack is explained in Section \ref{sec:logjam}. Different countermeasures are presented in Section \ref{sec:countermeasures}, and the main points of this paper are summarized in the conclusion in Section \ref{sec:conclusion}. 

The Logjam attack is applicable against TLS versions 1.0, 1.1 and 1.2. In this paper, when the term TLS is used, version 1.2 is meant except if stated otherwise.

\section{Breaking Diffie Hellman}
\label{sec:diffiehellman}
The Diffie-Hellman key exchange\cite{diffiehellman} is a cryptographic algorithm that is used in order to securely generate a shared secret between two communicating parties. This key can then be used as a symmetric key in order to allow the parties to communicate securely. 

\[
\begin{array}{@{} l c r @{}}
$Alice$ 				&  				& $Bob$ \\
\\
g, p					&				& g, p\\
$Generate $a \in \Z_p\\
						& \XR{g^a} \\
						&				& $Generate $b \in \Z_p\\
						& \XL{g^b} \\
s \leftarrow (g^b)^a	&				& s \leftarrow (g^a)^b\\
 \end{array}
\]

After running the Diffie-Hellman key exchange protocol, Alice and Bob have obtained shared secret $s = g^{ab}$. An attacker that intercepts their communication will have the values of $g^a$ and $g^b$, but will be unable to obtain the shared secret $g^{ab}$. In order to obtain this, the attacker will have to compute a discrete logarithm to either obtain the secret exponent $a$ or $b$. The security of the Diffie-Hellman protocol resides in the fact that solving the discrete log problem is computationally hard. 

Solving the Discrete Log problem for large groups $p$ is efficiently done using the number field sieve algorithm\cite{schirokauer2008impact}. Other algorithms exist that can solve the discrete log problem faster, such as the one presented by Barbulescu et al. \cite{barbulescu2014heuristic}, but these algorithms have additional constraints\footnote{The algorithm by Barbulescu et al. requires the finite field to be of small characteristic}, rendering them unsuitable for attacking real-world Diffie-Hellman key exchanges. 

The Number Field Sieve algorithm consists of four steps. We will not discuss the algorithm in detail, but will present a short overview of the steps, as this is relevant for understanding the Logjam attack. More details about the number field sieve algorithm can easily be found online, a great introduction is presented in \emph{A Tale of Two Sieves}\cite{pomerance2008tale} by Pomerance. 

The first step is Polynomial selection, which takes a relatively short time and is easily parallelized, but is important as it has great influence on the running time of the second step\cite{bai2011polynomial}, which is the sieving step. During the sieving, ranges of field elements are factored in order to find elements that are B-smooth\footnote{A number is B-smooth if it only consists of prime factors less than B. In other words, it can be factored in relatively small primes. }. Sieving also parallelizes well. The third step is linear algebra. A large matrix is built, consisting of the coefficient vectors found in the previous step. This step can, to some extent, also be parallelized. The fourth and final step is the descent. In this step, the output of the previous step is used to compute the discrete log for a given $a, g$. 

\section{TLS}
\label{sec:tls}
Before we discuss how Logjam uses the number field sieve algorithm in order to attack TLS, we first discuss in some detail how the relevant part of TLS works\footnote{Many ways of using TLS are possible, like using client certificates for authentication. We do not discuss these features and limit ourselves to the TLS handshake protocol in the most common scenario, where the only the server authenticates to the user by means of a signed certificate. }. 

When a client connects to a server through TLS, he sends a \texttt{ClientHello} message. The format of the message is as follows, as specified\footnote{For the sake of brevity, for some of the TLS messages mentioned in this section, the full message structure definition is omitted or shortened.} in RFC 5246\cite{rfc5246}:
\begin{verbatim}
  struct {
      ProtocolVersion client_version;
      Random random;
      SessionID session_id;
      CipherSuite cipher_suites<2..2^16-2>;
      CompressionMethod compression_methods<1..2^8-1>;
      select (extensions_present) {
          case false: struct {};
          case true: Extension extensions<0..2^16-1>;
      };
  } ClientHello;
\end{verbatim}
      
The server receives the \texttt{ClientHello} message, and picks a cipher from the CipherSuite list that both client and server support\footnote{Ideally, the server picks the most secure cipher suite that is mutually supported.}. For the Logjam attack, the DHE ciphersuites are relevant. The server replies with a \texttt{ServerHello} message, structured as follows:

\begin{verbatim}
  struct {
      ProtocolVersion server_version;
      Random random;
      SessionID session_id;
      CipherSuite cipher_suite;
      CompressionMethod compression_method;
      select (extensions_present) {
          case false: struct {};
          case true: Extension extensions<0..2^16-1>;
      };
  } ServerHello;
\end{verbatim}

The \texttt{ServerHello} is followed by a \texttt{ServerCertificate} message, containing containing the server's certificate and the certificate chain linking the certificate to a trusted Certificate Authority. In the case of a DHE-family ciphersuite, the server now sends a \texttt{ServerKeyExchange} message

\begin{verbatim}
  struct {
      select (KeyExchangeAlgorithm) {
          case dh_anon:
              ServerDHParams params;
          case dhe_dss:
          case dhe_rsa:
              ServerDHParams params;
              digitally-signed struct {
                  opaque client_random[32];
                  opaque server_random[32];
                  ServerDHParams params;
              } signed_params;
  } ServerKeyExchange;
\end{verbatim}

Note that the \texttt{ServerKeyExchange} contains signed Diffie-Hellman parameters, but does not contain the actual cipher suite that was chosen by the server. This is relevant for the Logjam attack, as explained in the next section. The server ends the Hello phase by sending a \texttt{ServerHelloDone} message. The client has now received the server's Diffie-Hellman parameters, namely the modulus $p$, generator $g$ and the server's Diffie-Hellman public value $Y_s$, and responds to the server by sending a \texttt{ClientKeyExchange} message, containing the client's Diffie-Hellman public value $Y_c$. The client now sends the \texttt{Finished} message, securely encrypted with the negotiated cipher and key. The server responds by also sending a \texttt{Finished} message, finalizing the handshake. 

Clients \emph{may} use the so-called \emph{False Start}\cite{langley2010transport}. This allows the client send an encrypted data packet \emph{before} it received the Finished message from the server, in order to reduce the latency caused by the TLS handshake protocol. 

\section{Logjam}
\label{sec:logjam}
Logjam\cite{adrianimperfect} is an attack on TLS, that makes use of a protocol flaw in TLS in conjunction with support for obsolete and insecure protocols. The attack works as follows: 

\subsection{Target selection}
In order to be vulnerable for the Logjam attack, there are some properties that must hold for the server configuration. The server should either support an export-grade cipher suite, or support DHE with a small prime. Export-grade ciphers are ciphers that were intentionally weakened in the 1990s in order to comply with US export laws on cryptographic technology. 

The authors of the original research paper demonstrated that attacking 512-bit primes is very possible with today's technology, while the precomputation for 768-bit primes should be feasible for academic research teams. According to Adrian et al., it is possible that nation-state level opponents could break a small number of 1024-bit Diffie-Hellman groups, although the precomputation would still take a very long time. They argue that this is possibly already the case, based on evidence of the NSA's VPN decryption infrastructure that were leaked by Edward Snowden\cite{spiegel_snowden}.

An important issue is that some primes are used by many different servers. This allows an attacker to do the precomputation once, and use the log db in order to attack many different servers. While computation for 512-bits primes is not extremely expensive, this definitely increases the plausibility of attacks against 1024-bit Diffie-Hellman groups, as this would only be sufficiently rewarding if more than a single target server can be attacked using the log db. According to Adrian et al., at the time they published their research paper, 8.4\% of the top million websites are vulnerable for the Logjam attack as they support the use of 512-bit groups. An attacker who is able to compute a log db for the \emph{single most used 1024-bit group} is able to attack 37.1\% of the one million most popular websites. 

\subsection{Precomputation of log db}
Once the prime $p$ that is used by the targeted server(s) is known, the attacker proceeds to perform a precomputation. This largely follows the steps outlined in Section \ref{sec:diffiehellman}, but the precomputation only involves the first three steps. The reason is that the first three stages of the algorithm are solely based on the prime $p$. If we want to compute $a$ from $y = g^a \mod p$, we only need to use the values of $y$ and $g$ in the descent stage. Logjam exploits this by precomputing the first three steps for a desired prime $p$, and storing the intermediate result in a \emph{log db}. When attempting to compute a discrete log in the field $\Z_p$, the log db is used as input for the descent phase. 

Computing a log db takes a long time. However, it is important to know that, while the Logjam attack focuses on downgrading a TLS connection to export grade crypto (as explained in the next step), a nation-state level attacker could possibly attack servers that only support regular, 1024-bit DHE cipher suites. 

\subsection{Man-in-the-Middle of a connection}
\begin{figure}[!ht]
  \centering
  \includegraphics[width=1\textwidth]{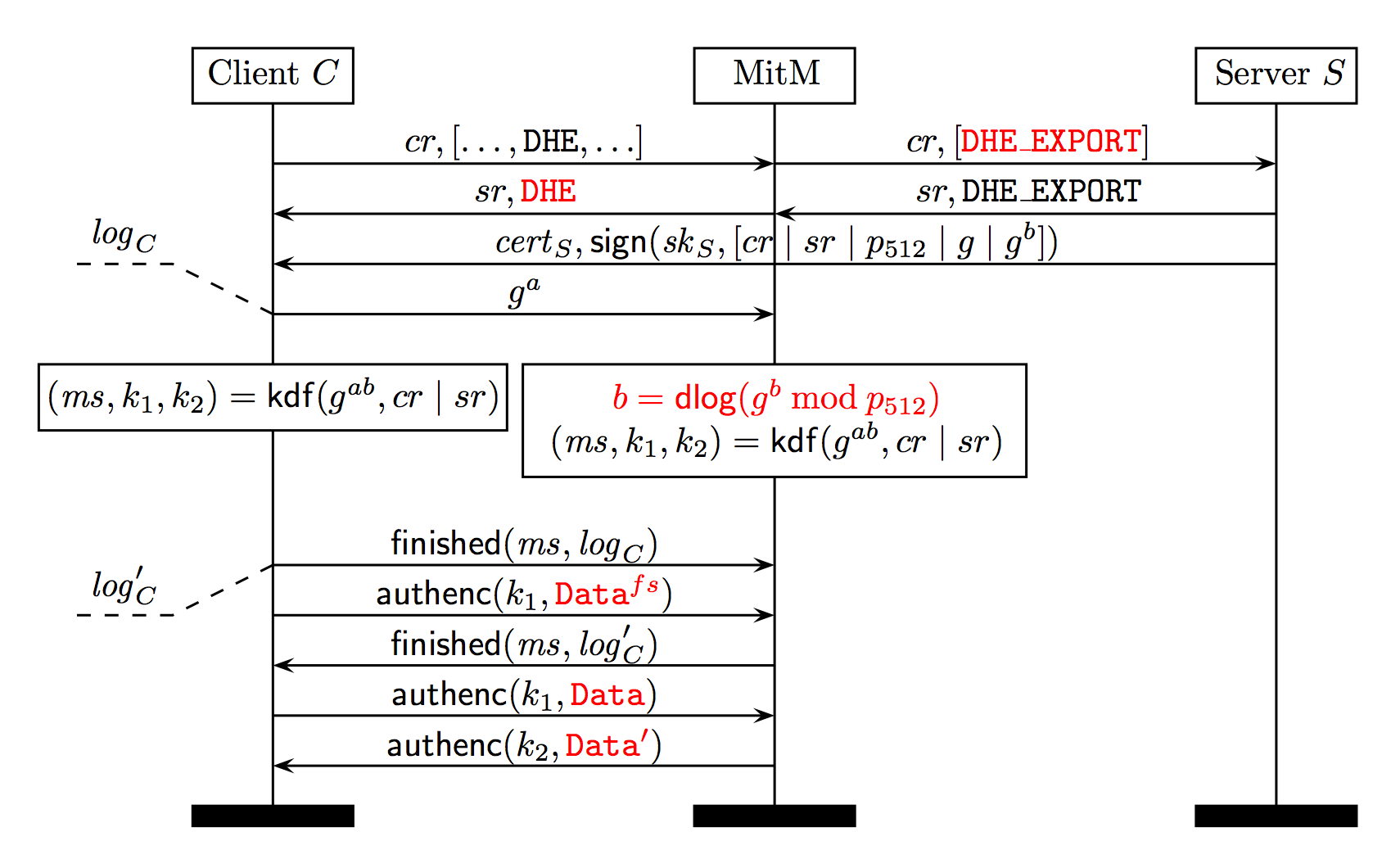}
  \caption{Message Sequence Chart of the Logjam attack. Source: \cite{adrianimperfect}}
  \label{fig:logjam_msc}
\end{figure}

When the attacker has built a log db for a given prime, he can abuse a Man-in-the-Middle scenario and intercept and decrypt the communication between the client and the server. This works as illustrated in figure \ref{fig:logjam_msc}.

The \texttt{ClientHello} message is modified by the attacker, removing all cryptographic suites and adding an export-grade suite that is supported by the server. The server trivially accepts to use this suite, as it is the only suite they both 'support'. The \texttt{ServerHello} message that is sent to the client is now modified, the DHE\_EXPORT cipher choice is replaced by regular DHE as not to arouse suspicion and to not depend on DHE\_EXPORT support on the client side. When the server sends his signed \texttt{ServerKeyExchange} message, this discrepancy between cipher suite choices between client and server is not a problem, as the DHE parameters have the same format. The server also sent his Diffie-Hellman public value $g^b$, and the attacker starts to solve the discrete log problem using the log db, in order to uncover $b$. The client will send his public value, and send the \texttt{Finished} message. As soon as the attacker found the value $b$, the attacker sends the message and herewith completes the TLS handshake. He is now able to decrypt any encrypted traffic between client and server, and can also modify and re-encrypt data. 

If the client uses False Start, he might send a first encrypted data packet before receiving the server's \texttt{Finished} message. In that case, the attacker may decrypt the packet at any time when the discrete log has been found. Even with the log db, finding $b$ is computationally hard, but there are tricks to hide or increase the time-frame required by the attacker in order to be able to send a valid Finished message. The client's browser will timeout if the server does not send any new TLS message within a certain time-frame. However, the attacker can send a TLS warning alert which, besides resetting the timeout, is ignored by the client. The attacker could open a TLS session to a server the client is \emph{expected} to connect to by injecting an invisible element from the target server in an unencrypted webpage that is also visited by the client, and solve the discrete log with little time pressure. This connection can then be kept alive, in order to avoid suspicion due to a long wait when the client visits the target server.

\section{Countermeasures}
\label{sec:countermeasures}
Logjam relies on several server and client settings, and different possibilities exist to limit or fully mitigate the risks of the Logjam attack. In this section, we discuss several safeguards that may be implemented by the server and/or by the client. 

\subsection{Server-side countermeasures}
\begin{itemize}
	\item Randomization of the parameter $p$. Although generating a prime $p$ that is safe to use\footnote{Not all primes are safe, there are some constraints on which primes are suitable for use. Some servers actually use unsafe primes. David et al. \cite{adrianimperfect} found 6.8\% of servers are configured with a $p$ for which $(p − 1)/2$ is composite, potentially rendering them vulnerable to attack if it consists solely of relatively small prime factors\cite{pohlig1976improved}.} is computationally expensive, it should be possible to let the server generate one upon installation. In that case, there is less reward for an attacker to do the precomputation for any prime $p$, as it will allow him to attack only a single server. Additionally, the prime could be regenerated periodically, in order to further limit the potential gain for an attacker. 

	\item Disable support for export-grade cipher suites. Export grade cipher suites use small, 512-bits primes, and as such must be considered unsafe. The server should not be using primes shorter than 2048 bits. 
	
	\item Use ECDHE instead of DHE. ECDHE is a variant of Diffie-Hellman which operates in an elliptic curve instead of a finite field. The advantage here is that discrete log algorithms for elliptic curves do not benefit as much from precomputation, raising the bar for an attacker. 
\end{itemize}
	
\subsection{Client-side countermeasures}
\begin{itemize}
	\item Require long primes $p$. Currently, 2048 bits primes are considered secure, while 1024 bits primes may be compromised when considering nation-state level opponents. Note that the client should actually check if $p > 2^{1024}$ as a 1024 bits prime can simply be padded with zeroes in order to be accepted as a 2048 bits.

	\item Reduce TLS handshake timeout. This provides the attacker with a shorter time window in which he must solve the DL, increasing the difficulty of the attack. Also, the timeout should be separate from the current timer, in order to ensure that it cannot be reset by the attacker using TLS warning alert messages. 

	\item Be careful with TLS False Start. Even when using shorter timeouts (as discussed before), the data that is transmitted before the handshake completes can be decrypted by an attacker as soon as he solved the discrete log for this session and obtains the shared secret. 
\end{itemize}

\section{Conclusion}
\label{sec:conclusion}
It is expected that TLS 1.3 will mitigate the Logjam attack by adding a field to the \texttt{ServerKeyExchange} message, that contains the cipher suite chosen by the server. This field should, like the Diffie-Hellman parameters, be protected against tampering by a signature. This solves the vulnerability, as an attacker can no longer trick the client into believing a strong cipher suite was chosen, while the server chose a deprecated and insecure cipher suite. TLS 1.2, however, can be expected to remain supported for a long time, and until support is phased out the Logjam attack remains a serious threat. Precautions must be taken in order to limit the risks of abuse, and different countermeasures exist. In the opinion of the author, a solution that is in most cases relatively easy to implement would be to either adopt strong, 2048-bit primes, or to diversify 1024-bit primes by (preferably periodically) generating them on the server. This would either make precomputation infeasible or at least drastically decrease the advantage of precomputation, as the resulting log db could only be used against a single server. Other solutions, like transitioning away from DHE towards ECDHE, are possibly more secure, but require larger changes to existing infrastructure, as not all software supports the stronger ECDHE cipher suite families.

\bibliographystyle{plain}
\bibliography{citations.bib}

\begin{thebibliography}{10}

\bibitem{adrianimperfect}
David Adrian, Karthikeyan Bhargavan, Zakir Durumeric, Pierrick Gaudry, Matthew
  Green, J~Alex Halderman, Nadia Heninger, Drew Springall, Emmanuel Thom{\'e},
  Luke Valenta, et~al.
\newblock Imperfect forward secrecy: How diffie-hellman fails in practice.
\newblock 2015.

\bibitem{al2013lucky}
Nadhem~J Al~Fardan and Kenneth~G Paterson.
\newblock Lucky thirteen: Breaking the tls and dtls record protocols.
\newblock In {\em Security and Privacy (SP), 2013 IEEE Symposium on}, pages
  526--540. IEEE, 2013.

\bibitem{bai2011polynomial}
Shi Bai.
\newblock {\em Polynomial selection for the number field sieve}.
\newblock PhD thesis, PhD thesis, Australian National University, 2011.

\bibitem{barbulescu2014heuristic}
Razvan Barbulescu, Pierrick Gaudry, Antoine Joux, and Emmanuel Thom{\'e}.
\newblock A heuristic quasi-polynomial algorithm for discrete logarithm in
  finite fields of small characteristic.
\newblock In {\em Advances in Cryptology--Eurocrypt 2014}, pages 1--16.
  Springer, 2014.

\bibitem{beurdouche2015messy}
Benjamin Beurdouche, Karthikeyan Bhargavan, Antoine Delignat-Lavaud, C{\'e}dric
  Fournet, Markulf Kohlweiss, Alfredo Pironti, Pierre-Yves Strub, and
  Jean~Karim Zinzindohoue.
\newblock A messy state of the union: Taming the composite state machines of
  tls.
\newblock In {\em IEEE Symposium on Security and Privacy}, 2015.

\bibitem{rfc5246}
Tim Dierks and Eric Rescorla.
\newblock The transport layer security (tls) protocol version 1.2, august 2008.
  url http://www. ietf. org/rfc/rfc5246. txt.
\newblock Technical report, RFC 5246, 2008.

\bibitem{diffiehellman}
Whitfield Diffie and Martin~E Hellman.
\newblock New directions in cryptography.
\newblock {\em Information Theory, IEEE Transactions on}, 22(6):644--654, 1976.

\bibitem{letsencrypt}
Linux Foundation.
\newblock About let's encrypt.
\newblock \url{https://letsencrypt.org/about/}.
\newblock Accessed: 27-01-2016.

\bibitem{langley2010transport}
A~Langley, N~Modadugu, and B~Moeller.
\newblock Transport layer security (tls) false start.
\newblock {\em draft-bmoeller-tls-falsestart-00, June}, 2, 2010.

\bibitem{smacktls}
Inria Microsoft.
\newblock Smack: State machine attacks.
\newblock \url{https://mitls.org/pages/attacks/SMACK}.
\newblock Accessed: 27-01-2016.

\bibitem{pohlig1976improved}
Stephen~C Pohlig and Martin~E Hellman.
\newblock An improved algorithm for computing logarithms over gf (p) and its
  cryptographic significance. 1978 january: Ieee transactions on information
  theory.
\newblock {\em Article submitted on}, 1976.

\bibitem{pomerance2008tale}
Carl Pomerance.
\newblock A tale of two sieves.
\newblock {\em Biscuits of Number Theory}, 85, 2008.

\bibitem{schirokauer2008impact}
Oliver Schirokauer.
\newblock The impact of the number field sieve on the discrete logarithm
  problem in finite fields.
\newblock In {\em Proceedings of the 2002 Algorithmic Number Theory workshop at
  MSRI}, 2008.

\bibitem{spiegel_snowden}
Der Spiegel.
\newblock Prying eyes: Inside the nsa's war on internet security.
\newblock
  \url{http://www.spiegel.de/international/germany/inside-the-nsa-s-war-on-internet-security-a-1010361.html}.
\newblock Accessed: 27-01-2016.

\end{thebibliography}

\end{document}